\documentclass[%
 reprint,
superscriptaddress,
 amsmath,amssymb,
 aps,
 nofootinbib
]{revtex4-1}

\def\be#1{\begin{equation}\label{#1}}
\def\ee{\end{equation}}
\newcommand {\ba}[2]{\be{#1}\begin{array}{#2}}
	\newcommand {\ea}{\end{array} \ee}
\def\eq#1{(\ref{#1})}
\newcommand {\hence}{\quad\Rightarrow\quad}
\newcommand{\mr}{\mathrm}

\newcommand{\qq}{\,,\qquad}

\newcommand\Top{\rule{0pt}{2.8ex}}       
\newcommand\Bot{\rule[-1.2ex]{0pt}{0pt}} 

\def\({\left(}
\def\){\right)}

\def\kB{k_{\!B}}

\let\ka=k

\usepackage{ifpdf}

\newlength{\pix}
\AtBeginDocument{%
	\setlength{\oddsidemargin}{\dimexpr(\paperwidth-\textwidth)/2-1in}%
	\setlength{\evensidemargin}{\oddsidemargin}%
	\setlength{\topmargin}{%
		\dimexpr(\paperheight-\textheight)/2-\headheight-\headsep-1in}%
}
\usepackage{floatrow}
\usepackage[justification=centering]{caption}

\usepackage{graphicx}
\usepackage{dcolumn}
\usepackage{bm}
\usepackage{xfrac}

\usepackage{cleveref}
\usepackage{stackrel}

\begin{document}

\preprint{APS/123-QED}

\title{Change of entropy for the one-dimensional ballistic heat equation: Sinusoidal initial perturbation}
\author{Aleksei A. Sokolov}
\email{sokolovalexey1@gmail.com} 
\affiliation{Continuum Mechanics and Materials Theory, Technische Universit\"at Berlin, Einsteinufer 5, 10587 Berlin, Germany}
\affiliation{Theoretical and Applied Mechanics, Peter the Great Saint Petersburg Polytechnic University, Politekhnicheskaja 29, 195251 Saint Petersburg, Russia}
\author{Anton M. Krivtsov}
\email{akrivtsov@bk.ru}
\affiliation{Theoretical and Applied Mechanics, Peter the Great Saint Petersburg Polytechnic University, Politekhnicheskaja 29, 195251 Saint Petersburg, Russia}
\affiliation{Institute for Problems in Mechanical Engineering of the Russian Academy of Sciences, Bol'shoy pr. 61, V.O., 199178 Saint Petersburg, Russia}

\author{ Wolfgang H. M\"uller}
 \email{wolfgang.h.mueller@tu-berlin.de}
\affiliation{Continuum Mechanics and Materials Theory, Technische Universit\"at Berlin, Einsteinufer 5, 10587 Berlin, Germany}

\author{Elena N. Vilchevskaya}
\email{vilchevska@gmail.com}
\affiliation{Theoretical and Applied Mechanics, Peter the Great Saint Petersburg Polytechnic University, Politekhnicheskaja 29, 195251 Saint Petersburg, Russia}
\affiliation{Institute for Problems in Mechanical Engineering of the Russian Academy of Sciences, Bol'shoy pr. 61, V.O., 199178 Saint Petersburg, Russia}

\date{received 24 August 2018; published 4 April 2019}

\begin{abstract}
This work presents a thermodynamic analysis of the ballistic heat equation from two viewpoints: classical irreversible thermodynamics (CIT) and extended irreversible thermodynamics (EIT). A formula for calculating the entropy within the framework of EIT for the ballistic heat equation is derived. The entropy is calculated for a sinusoidal initial temperature perturbation by using both approaches. The results obtained from CIT show that the entropy is a non-monotonic function and that the entropy production can be negative. The results obtained for EIT show that the entropy is a monotonic function and that the entropy production is nonnegative. A comparison between the entropy behaviors predicted for the ballistic, for the ordinary Fourier-based, and for the hyperbolic heat equation is made. A crucial difference of the asymptotic behavior of the entropy for the ballistic heat equation is shown. It is argued that mathematical time reversibility of the partial differential ballistic heat equation is not consistent with its physical irreversibility. The processes described by the ballistic heat equation are irreversible because of the entropy increase.

\end{abstract}

\pacs{Valid PACS appear here}
\maketitle


\section{Introduction}
Classical thermodynamic approaches lead to the Fourier based parabolic law of heat conduction. The propagation of heat described by the classical heat conduction equation is observed on the macroscale and frequently used in engineering applications. However, considering the problem of heat conduction from the atomistic viewpoint can lead to different results. Various models of lattices (considering anharmonicity of the interatomic bonds, 2D lattices) are frequently used for the description of heat transfer \cite{Dhar_and_Dandekar_2015,Savin_and_Kosevich_2014,Babenkov_et_al._2016,Saadatmand_et_al.2018,Lepri_et_al._2003,Gavrilov_2018,GOLDSTEIN2007235,Guzev2007,Zheng2011,Falasco2015, Dhar2008, Savin2007, Landi2013,Zhou2016,Kannan2012}. Modern technology demonstrated experimentally \cite{Chang_2008} that Fourier's law is violated in low-dimensional nanostructures, where a ballistic type of heat conduction is observed. This motivates recent interest in properties of structures, such as graphene and carbon nanotubes \cite{Porubov_and_Berinsky_2014,Berinsky_et_al._2014} and, in particular, in their thermal properties \cite{Faugeras_et_al._2010}. From a theoretical point of view one of the most attractive playgrounds for the investigation of heat conduction is the harmonic one-dimensional crystal, since all of its thermodynamic properties can be obtained analytically from the equations of lattice dynamics. A pioneering contribution regarding the the one-dimensional crystal was made by Schr\"odinger \cite{Shroedinger_1914}. He obtained an exact solution for the displacements with arbitrary initial conditions in terms of Bessel functions. This work became a foundation for a future investigation of energy transfer in a one-dimensional chain by Hemmer \cite{Hemmer_1959}.
A new approach for the description of nonequilibrium heat conduction processes in crystals was developed in Refs. \cite{Krivtsov_2015,Krivtsov_2014,Kuzkin_and_Krivtsov_2017}. A hyperbolic equation called ``ballistic heat equation'' was obtained as a mathematical consequence of the equations of lattice dynamics. From an experimental point of view such processes can be observed in low dimensional structures exposed to a laser excitation \cite{Indeitsev2017}.

The mathematical properties of the ballistic heat equation were investigated in several papers including Ref. \cite{Sokolov_2016}. The ballistic heat equation is reversible with respect to a substitution of $t$ to $-t$. However, it seems intuitively probable that processes described by the ballistic heat equation are irreversible. The calculation of the entropy production associated with processes described by the ballistic heat equation will help to reveal its thermodynamic properties and answer the question of reversibility. Pioneering ideas considering irreversible thermodynamical processes were mainly developed by Prigogine and Onsager \cite{Prigogine_1961,Onsager_1931}. Their investigations led to the formulation of classical irreversible thermodynamics (CIT). However, further research showed that CIT could not describe a wide class of phenomena including short time and small space scales, as well as hyperbolic models of heat conduction (Cattaneo type) allowing entropy production to be negative in some cases \cite{Jou_et_al._2010}. The phenomenological investigation \cite{Krivtsov_et_al._2017} of the ballistic heat equation showed indeed that within the framework of CIT the entropy production of processes described by the ballistic heat equation can be negative. 

One of the modifications of existing theories was extended irreversible thermodynamics (EIT) introduced in~\cite{Mueller_1966} and explained in details in Ref. \cite{Mueller_1993}. It was shown in Refs. \cite{Jou_et_al._2010,Mueller_1966,Criado-Sancho_1993,Alvarez2008} that this theory is applicable for more complex models of heat conduction, such as the hyperbolic heat equation (Cattaneo type).

The aim of this work is to consider the ballistic heat equation within the framework of EIT, to obtain general formulas, which allow us to calculate the entropy for this model, to consider the particular problem of a sinusoidal initial temperature perturbation, and to compare the results obtained by CIT and EIT for the ballistic heat equation and hyperbolic and classical heat conduction equations.

\section{Models of heat conduction}
If the heat capacity at a constant volume $c_V$ does not depend on time, the energy balance equation reads:
\be{enbal}
 \rho c_V \dot{T} = - h'
,\ee
where $\rho$ is the density, $T$ is the temperature and $h$ is the heat flux, and the dot $()\dot{}$ and the dash $()'$ denote temporal and spatial derivatives, respectively. In this paper we are going to analyze three models of heat conduction. 
\paragraph{The classical heat equation:}
\be{f1}
\dot T = \alpha T'',
\ee
where $\alpha = \sfrac{\kappa}{\rho c_V}$ is the coefficient of thermal diffusivity, $\kappa$ is the coefficient of thermal conductivity. Equation \eq{f1} is obtained on the basis of Fourier's law:
\be{ff}
h = - \kappa T'.
\ee
\paragraph{Hyperbolic equation (Maxwell-Cattaneo-Vernotte type):}
\be{mcv1}
\ddot{T} + \frac{1}{\tau}\dot{T} = \frac {\alpha}{\tau} T'',
\ee
where $\tau$ is the relaxation time.
It was obtained by introducing a heat flux relaxation term to Eq. \eq{ff} \cite{Jou_et_al._2010}:
\be{fmcv}
\dot{h} + \frac{1}{\tau} h = - \frac{\kappa}{\tau} T'.
\ee

\paragraph{The ballistic heat equation~\cite{Krivtsov_2015}:}
\be{4}
\ddot T + \frac1 t\,\dot T = c^2 T'',
\ee
where $c$ is the speed of sound, $t$ is the time passed from the moment of the instantaneous heat perturbation. The corresponding equation for the heat flux reads:
\be{fb}
	\dot h + \frac1 t\,h = -\rho \kB c^2 T'.
\ee
Equation \eq{4} was first derived in Ref. \cite{Krivtsov_2015} from the equations of lattice dynamics of a one-dimensional crystal with nearest neighbor linear force interaction: 
\be{b1}
\ddot{q}_i = \omega_e^2( q_{i-1} - 2 q_i + q_{i+1}), \quad \omega_e \stackrel{\mathrm{def}}{=} \sqrt{ \frac{C}{m}},
\ee
where $q_i$ is the displacement of the particle with index~$i$, $C$ is the interatomic bond stiffness, and $m$ is the particle mass.

The following initial conditions are considered:
\be{b2}
q_i|_{t = 0} = 0, \qquad \dot{q}_i|_{t=0} = \sigma(x) \rho_i,
\ee
where $\rho_i$ are independent random variables with zero expectation and unit variance; $\sigma$ is the variance of the initial particle velocity. The variance is a slowly changing function of the spatial coordinate $x = ia$, where $a$ is the initial distance between neighboring particles.

The following definitions of the kinetic temperature and the heat flux were used for the derivation:
\be{b3}
k_BT = m  \langle \dot{q_i} \rangle^2, \quad h = \frac{1}{2}C \langle (q_i - q_{i+1})(\dot{q}_i - \dot{q}_{i+1})  \rangle,
\ee
where $\langle \rangle$ denotes the expectation value. According to the Dulong-Petit law for a one-dimensional system, Boltzmann's coefficient $\kB$ at the right hand side of Eq. \eq{fb} is the specific heat capacity of the one-dimensional crystal at a constant volume $c_V$.

The classical heat Eq. \eq{f1} is solved in combination with the following initial condition:
\be{fi}
T|_{t=0} = \theta_0(x).
\ee
Equations \eq{mcv1} and \eq{4} are to be solved by using the initial conditions:
\be{b5}
T|_{t=0} = \theta_0(x), \qquad h|_{t=0} = 0.
\ee
According to Eq. \eq{enbal} the above condition for the heat flux is equivalent to:
\be{b5a}
\dot{T}|_{t=0} = 0.
\ee

The hyperbolic Eq. \eqref{mcv1} and the ballistic heat Eq.~\eq{4} have similar forms and a somewhat similar behavior (e.g., a finite velocity of the heat front propagation). For the ballistic heat Eq. \eq{4} the speed of heat propagation is the speed of sound of the medium, $c = \omega_e a$ \cite{Krivtsov_2015}. For the hyperbolic heat Eq.~\eq{mcv1} the speed of heat propagation is $c^2_{\mathrm{MCV}}= \alpha/\tau$ \cite{ziik1984}.

However, there is a significant difference between these two equations: the material constant $\tau$ is replaced in the ballistic heat equation by the physical time, $t$ \cite{Krivtsov_2015}.

From the form of Eq. \eq{4} it seems that it has a singularity. However, when Eq. \eq{4} is solved together with the initial conditions Eqs. \eq{b5} and \eq{b5a}, the singularity is absent, which is confirmed by the general analytical solution \cite{Krivtsov_Kuzkin_2018}:
\be{bs}
T(x,t) = \frac{1}{2\pi} \int_0^{2 \pi} \theta_0 \(x + ct \cos p\) dp
\ee
and solutions of the particular initial problems \cite{Sokolov_2016}.

We note that the exact solution for the atomic velocities and displacements (see Refs. \cite{Shroedinger_1914,Hemmer_1959}) of the Eq.~\eq{b1} predicts an infinite speed of signal propagation, while the ballistic equation \eq{4} describes a propagation of temperature at a finite speed, $c$. This difference is due to the fact that Eq. \eq{4} is obtained from Eq. \eq{b1} by using continualization and coarse graining in space \cite{Krivtsov_2015}.

Also note that Eq. \eq{4} describes ballistic heat conduction and does not describe a transition from ballistic to diffusive regimes. It is known that the ballistic heat transport occurs when phonons can propagate without scattering. It happens when the size of the system is comparable to the mean-free path of the carriers. In this case the thermal conductivity is size dependent. However, in a harmonic one-dimensional chain the phonon mean free path is infinite~\cite{Jou_et_al._2010} and no phonon-phonon, phonon-impurity, or phonon-boundary scattering occurs. Thus the ballistic heat Eq. \eq{4} has no size effect and contains only one parameter---the speed of sound in the medium $c$. 

In this paper we will consider and compare the entropy for the models described above, namely: Fourier's heat equation, a phenomenological model which describes heat conduction at the macroscale; the hyperbolic heat equation, a modification of the previous one taking into account wave properties of the heat propagation; and the ballistic heat equation, which is obtained as a direct consequence of the lattice dynamics and is fully based on the wave processes in the crystal lattice.
\section {Entropy inequality for the ballistic heat equation}
The formalism of CIT is based on the hypothesis of local equilibrium. It postulates \cite{Jou_et_al._2010} that a thermodynamic system can be divided into a number of microscopic cells, each of which can be treated like a macroscopic system in equilibrium. In each cell the state variables remain uniform but they can change from cell to cell \cite{DeGroot_and_Mazur_1962}. They can also change with time so that they finally depend continuously on space and time coordinates, $(x, t)$ \cite{Jou_et_al._2010}. Following Ref. \cite{Krivtsov_et_al._2017} we consider in this work thermal perturbations only without the presence of mechanical motion. It means that the set of state variables is narrowed down to the specific internal energy, $u$, only. That leads us to the following form of a {\em Gibbs relation} \cite{Jou_et_al._2010}:
\be{i1}
\mr{d}s = \frac{1}{T} \mr du, 
\ee
where $s$ is the specific entropy.
However, models taking the independent character of fluxes into account, turn out to be inconsistent with approaches of CIT \cite{Jou_et_al._2010,Krivtsov_et_al._2017}. EIT introduced a way to avoid contradictions by considering new state variables among the set of basic independent variables. Let us demonstrate this approach when applied to the ballistic heat Eq. \eq{4}. 
We assume that the entropy depends not only on the internal energy but also on the heat flux, $h$ (in particular the entropy depends on~$h^2$ because it is independent of the heat flux direction) \mbox{ \cite{Jou_et_al._2010,Mueller_1966}}:
\be{1}
s=s(u,h^2) \hence
\dot s = \frac{\partial s}{\partial u}\,\dot u + 2 \frac{\partial s}{\partial h^2}  \dot h  h
.\ee
Let us write the Clausius-Duhem inequality in general form \cite{Jou_et_al._2010,Palmov_1998,Krivtsov_et_al._2017}:
\be{cd}
\rho (T\dot s - \dot u) - \frac{h T'}T \ge 0
,\ee
As shown in Ref. \cite{Krivtsov_et_al._2017} the substitution of the second relation in Eq. \eq{1} into the inequality Eq. \eq{cd} yields:
\be{2}
\rho \(T \frac{\partial s}{\partial u} - 1\) \dot u + h\(2 \rho T \frac{\partial s}{\partial h^2}  \dot h  - \frac{ T'}T\) \ge 0
,\ee
It is assumed in Refs. \cite{Colemann_1964,Jou_et_al._2010} that arbitrary energy supplies keep the balance of energy satisfied. Thus, the balance law does not impose constraints on $\dot{u}$. Therefore, in Eq. \eq{2} $\dot{u}$ can take arbitrary independent values. Thus, to guarantee that Eq. \eq{2} is satisfied, it follows that:
\be{3}
\frac{\partial s}{\partial u} = \frac{1}{T} \qq
h\(2 \rho T^2 \frac{\partial s}{\partial h^2}  \dot h  - T'\) \ge 0
.\ee
The relations Eqs. \eq{3} were discussed in previous work \cite{Krivtsov_et_al._2017}.

Let us now consider the second inequality in Eq. \eq {3} in context with the equation of ballistic thermal conductivity Eq. \eq {4}. By expressing $ T '$ with the first equation from Eq. \eq {fb} and then substituting it into the inequality from Eq. \eq {3} we obtain:
\be{5}
\dot{h}h \( 2\rho^2 \kB T^2 c^2  \frac{\partial s}{\partial h^2 } +  1\) + \frac{h^2}{t} \ge 0
.\ee
To keep this inequality satisfied for any values of $\dot{h}$, the first term must be zero. Therefore:
\be{6}
\frac{\partial s}{\partial h^2 } = - \frac{1}{2 \rho^2 c^2 c_V T^2 }
.\ee
By substituting the first relation from Eq. \eq{3} and the relation Eq. \eq{6} into Eq. \eq{1} yields:
\be{7}
\begin{aligned}
	&\dot{s} = \frac{1}{T} \dot{u} - \frac{1}{ \rho^2 c^2 c_V T^2 } h \dot{h} = \dot{s}_{\mr{eq}}(u) + \dot{s}_{\mr{ne}}(h), \\
	&\dot{s}_{\mr{eq}} = \frac{1}{T} \dot{u}, \quad
	\dot{s}_{\mr{ne}} = - \frac{1}{ \rho^2 c^2  c_V T^2} h \dot{h},
\end{aligned}
\ee
and as a differential, as an analogy to Eq. \eq{i1},
\be{7a}
	\mathrm{d} s = \frac{1}{T} \mathrm{d} u - \frac{1}{ \rho^2 c^2 c_V T^2 } h \mathrm{d}h,
\ee
where $s_{\mr{eq}}$ is the equilibrium part of the entropy change, which depends only on the specific internal energy, and $s_{\mr{ne}}$ is the nonequilibrium part, which is dependent on the heat flux. 

Thus, by considering additional parameters of state (heat flow), one can avoid contradictions leading to a violation of the second law when using the formulation of CIT. At the same time, the fact that heat can flow from cold to hot, which is observed for the ballistic heat Eq. \eq{4}, is not paradoxical, because it is caused by the inertia of the process under consideration.

Let us suppose that the temperature deviations are small. Then the heat capacity at constant volume, $c_V$, can be considered to be constant. By postulating further that the internal energy is a function of temperature we obtain:
\be{i2}
\mr du = c_V \mr dT.
\ee
By taking into account Eq. \eq{i2} the relations Eqs. \eq{i1} and \eq{7} have the following differential form:
\be{i3}
\dot{s}_{\mathrm{CIT}} = c_V\frac{\dot T}{T},
\ee
\be{i3a}
\dot{s}_{\mathrm{EIT}} = c_V\frac{\dot{T}}{T} - \frac{1}{ \rho^2 c^2 c_V T^2 } h \dot{h}.
\ee
\crefname{equation}{Eq.}{Eqs.} The functions describing the temperature $T(x,t)$ and the heat flux $h(x,t)$ are obtained as solutions of the \cref{f1,ff,mcv1,fmcv,4,fb}. Then by integrating Eqs. \eq{i3} and \eq{i3a} the corresponding entropies are found.

\section{Sinusoidal initial heat perturbation }
We now present an application of the relations for the entropy in the CIT approach Eq. \eq{i3} and in the EIT approach~Eq. \eq{i3a}. To this end consider a sinusoidal initial temperature distribution in an infinite adiabatic one-dimensional system of the form:
\be{init}
\theta_0(x) = \delta T \cos {k x} + T_0 
,\ee
where $ \delta T $ and $ T_0 $ are positive constants, which have the dimension of a temperature, and $ \ka $ is a wave number. Let us suppose that functions describing the temporal and spatial evolution of temperature and of the heat flux have the form:
\ba{8}{c}
T(t,x) = \delta T f_T(\omega t)\cos{k x} + T_0, \\
h(t,x) = \delta T c_V \rho c f_h(\omega t)\sin{k x}
,\ea
where $f_T$, $f_h$ are dimensionless functions of a dimensionless quantity, $\omega t$, where $\omega$~is a parameter with the dimension $1/\mathrm{s}$. To fulfill the initial conditions we put in Eq. \eq{8} $f_T(0) = 1$, $f_h(0)=0$. Since the perturbation and the solution are laterally periodic we can consider a lateral interval of one period. The wave length of the initial perturbation is $L = 2 \pi / k$. We consider the interval $x \in [-L/2, L/2]$. Note that $h(t,-L/2)=h(t,L/2)=0$, thus there is no heat flow into or out of the system.

In the case of an adiabatically closed system as described above, there is no entropy flux coming in or out of the system according to Ref. \cite{Jou_et_al._2010}. In this case the {\it rate} of total entropy is equal to the total entropy {\it production} in the system, which should be nonnegative. In this paper we consider an adiabatic system, because in this case a decrease in the total entropy indicates negative entropy production.
\subsection{Classical irreversible thermodynamics} We substitute the first equation from Eq. \eq{8} into Eq. \eq{i3} and then perform a series expansion by a small parameter $ \frac {\delta T}{T_0}$ (since the temperature deviations are small and $\delta T << T_0)$ up to terms of second order:
\ba{c1}{l l}
\dot{s} = & c_V  \tfrac {\delta T}{T_0} \dot{f_T}(\omega t) \cos(k x) \\
&- c_V  \(\tfrac {\delta T}{T_0}\)^2  \dot{f_T}(\omega t) f_T(\omega t)\cos ^2(k x) + O^3\( \tfrac {\delta T}{T_0} \)
.\ea
The total entropy {\it rate} for the considered interval is given by integration over the whole system:
\be{c2}
\dot{S}(t) = \int_{-L/2}^{L/2} \rho \dot{s} \,\mr dx = -  \frac{1}{2} c_V  \rho L \dot{f_T}(\omega t) f_T(\omega t) \(\frac {\delta T}{T_0}\)^2
.\ee
After integrating over time we obtain:
\be{c3}
\mr{\Delta} S_{\rm CIT}(t) = \frac{1}{4} c_V \rho L  \left(1-f^2_T(\omega t )\right) \(\frac {\delta T}{T_0}\)^2
,\ee
where $\mr{\Delta} S(t) = S(t) - S_0$ is the {\it change} of entropy and $S_0$~denotes the initial entropy of the system.

\subsection{Extended irreversible thermodynamics}
The expression for the entropy change Eq. ~eqref{i3a} obtained in EIT consists of two terms. One term is the equilibrium part, which is dependent only on the internal energy. It was already calculated above by using the CIT approach. Due to the linearity of integration we can now calculate the nonequilibrium part and then add both results to obtain the full entropy. By substituting the relations Eq.~\eq{8} into the relation for $s_{\mr{ne}}$ from Eq.~\eq{7} and by performing a series expansion with a small parameter $ \frac {\mr\delta T}{T_0}$ up to terms of second order we obtain:
\be{e1}
\dot{s}_{\rm ne} = - c_V  f_h( \omega t ) \dot{f}_h(\omega t ) \(\frac {\delta T}{T_0}\)^2 \sin^2(k x) + \mathcal{O}^3\( \frac {\delta T}{T_0} \)
.\ee
The total entropy rate for the considered interval is given by:
\be{e2}
\dot{S}_{\rm ne}(t) = \int_{-L/2}^{L/2} \rho \dot{s}\, \mr dx = -  \frac{1}{2} c_V  \rho L \dot{f_h}(\omega t) f_h(\omega t) \(\frac {\delta T}{T_0}\)^2
.\ee
After integrating over time we obtain:
\be{e3}
S_{\rm ne}(t)  = - \frac{1}{4} c_V \rho L  f^2_h(\omega t ) \(\frac {\delta T}{T_0}\)^2
.\ee
Here the initial nonequillibrium part of the entropy of the system is zero, since there are no fluxes at the beginning: $S_{\mr{ne}0} = 0$, so $\mr\Delta S_{\mr{ne}} = S_{\mr{ne}}(t) - S_{\mr{ne}0} = S_{\mr{ne}}(t)$. By using this result and Eqs. \eq{c3} and \eq{e3} we obtain a relation for the total change of entropy:
\be{e4}
\mr\Delta S_{\rm EIT} =  \frac{1}{4} c_V \rho L  \left[1 - f^2_T(\omega t ) - f^2_h(\omega t) \right] \(\frac {\delta T}{T_0}\)^2
.\ee
We proceed to specify the functions $f_T$ and $f_h$, which depend on the chosen model of heat conduction.

\section{Calculation of the entropy for different models of heat conduction}
\subsection{Ballistic heat equation}
In the case of the ballistic heat Eq.~\eq{4} the solutions for the temperature and for the heat flux have the following forms:
\ba{e5}{l}
T(t,x) = \delta T J_0(k c t)\cos{k x} + T_0, \\
h(t,x) = \delta T \kB\rho c J_1(k c t)\sin{k x}
,\ea
where $J_n(x)$ is the Bessel function of the first kind of order $n$. By applying Eqs. \eq{c3} and \eq{e4} we have $\omega = k c$, $f_T = J_0$, $f_h = J_1$:
\be{e7}
\begin{aligned}
	& \mr\Delta S^{\rm B}_{\rm CIT} = \frac{1}{4} c_V \rho L  \left[ 1 - J^2_0(k c t ) \right] \(\frac {\delta T}{T_0}\)^2, \\
	& \mr\Delta S^{\rm B}_{\rm EIT} = \frac{1}{4} c_V \rho L  \left[1 - J^2_0(k c t ) - J^2_1(k c t) \right] \(\frac {\delta T}{T_0}\)^2,
\end{aligned}
\ee
where the superscript $\rm B$ means that entropies corresponds to the ballistic heat Eq.~\eq{4}.

\subsection{Hyperbolic equation (Maxwell-Cattaneo-Vernotte)}
The entropy for the one-dimensional hyperbolic Eq.~\eq{mcv1}, was considered in detail in Ref.~\cite{Criado-Sancho_1993}. Following \cite{Criado-Sancho_1993} we represent the results obtained for the entropy pertinent to the hyperbolic heat Eq. \eq{mcv1} in order to compare them with the ballistic heat equation and the classical heat Eq. \eq{f1}.
The solution of Eqs. \eqref{mcv1} with the initial conditions Eq. \eqref{b5} for a sinusoidal initial perturbation Eq. \eq{init} will be the following (see Appendix):
\ba{mcv2}{l}
4\alpha \tau k^2 < 1: \\
T(t,x) = \delta T e^{-\frac{t}{2\tau}} \( \cosh\omega t + A \sinh\omega t \) \cos k x + T_0,  \\ 
h(t,x) =
2c_V \rho  \alpha k \delta T A e^{-\frac{t}{2 \tau}} \sinh \omega t \sin k x, \\

4\alpha \tau k^2 > 1: \\

T(t,x) = \delta T e^{-\frac{t}{2 \tau}}\( \cos\omega^*t + A^* \sin\omega^*t \) \cos k x + T_0, \\
h(t,x) = 2c_V \rho  \alpha k \delta T A^*  e^{-\frac{t}{2 \tau}} \sin\omega^*t \sin k x
,\ea
where $\omega = \frac{\sqrt{1-4\alpha \tau k^2}}{2 \tau}$, $A = \frac{1}{\sqrt{1-4\alpha \tau k^2}}$, $\omega^* = \frac{\sqrt{4\alpha \tau k^2-1}}{2 \tau}$, $A^* = \frac{1}{\sqrt{4\alpha \tau k^2-1}}$,  $\kappa = \rho c_V \alpha$ is taken into account.
The entropy change reads as follows [from \eqref{c3} and \eqref{e4})]:
\ba{mcv3}{l}

4\alpha \tau k^2 < 1: \\
\mr\Delta S^{\rm H}_{\mr{CIT}} = \frac{c_V \rho L}{4}  \left[1 - e^{-\frac{t}{\tau}} \( \cosh\omega t + A \sinh\omega t \)^2 \right] \(\frac {\delta T}{T_0}\)^2, \\
4\alpha \tau k^2 > 1: \\
\mr\Delta S^{\rm H}_{\mr{CIT}} = \frac{c_V \rho L}{4}  \left[ 1 - e^{-\frac{t}{\tau}} \(\cos\omega^*t + A^* \sin\omega^*t \)^2 \right] \(\frac {\delta T}{T_0}\)^2, 
\ea
where the superscript ${\rm H}$ means that entropies corresponds to the hyperbolic heat Eq. \eq{mcv1}. Recall that we use Eq. \eq{7} for calculating the entropy based on EIT was obtained for the ballistic heat Eq. \eq{4}. To calculate the entropy for the hyperbolic heat Eq. \eq{mcv1} we use a different formula originally obtained in Refs. \cite{Jou_et_al._2010,Mueller_1966}:
\be{mcv}
\dot{s} = \frac{1}{T} \dot{u} - \frac{\tau}{ \rho^2 c_V \alpha T^2} h \dot{h}.
\ee
This leads to the following relations for entropy:
\ba{mcv4}{l}
4\alpha t k^2 < 1: \\
\begin{aligned}
\mr\Delta S^{\rm H}_{\mr{EIT}}  = &  \frac{1}{4} c_V \rho L \{ 1 -   e^{-\frac{t}{\tau}} \left[ \( \cosh\omega t  + A \sinh\omega t \)^2  \right. \\

& \left. - 4 \alpha \tau k^2  A^2 \sinh^2\omega t \right]\}  \(\frac {\delta T}{T_0}\)^2 ,
\end{aligned}
 \\
4\alpha t k^2 > 1: \\
\begin{aligned}
\mr\Delta S^{\rm H}_{\mr{EIT}} = & \frac{1}{4} c_V \rho L \{ 1 - e^{\frac{t}{\tau}} \left[ \(\cos\omega^*t + A^* \sin\omega^* t \)^2  \right. \\
 & \left.- 4  \alpha \tau k^2 A^{*2} sin^2\omega^*t \right] \} \(\frac {\delta T}{T_0}\)^2.
\end{aligned}

\ea

\subsection{Fourier heat conduction equation.}
Let us now consider an application of Eq. \eq{c3} obtained  from CIT to the classical Fourier based heat Eq. \eq{f1}. The solution of Eq. \eq{f1} for a sinusoidal initial distribution has the form \cite{Carslaw_1959}:
\be{f_sol}
T(t,x) = \delta T e^{-\alpha k^2 t}\cos{k x} + T_0 
.\ee
Here we do not consider the heat flux, since CIT does not take it into account as state variable.
By applying Eq. \eq{c3} we find with $\omega = - \alpha k^2, f_T = e^{-\alpha k^2 t}$:
\be{f2}
\mr\Delta S^{\rm F}_{\mr{CIT}} = \frac{1}{4} c_V \rho L \( 1 - e^{-2 \alpha k^2 t}\) \(\frac {\delta T}{T_0}\)^2,
\ee
where the superscript $\rm F$ means that the entropy corresponds to the Fourier-based classical heat conduction Eq.~\eq{f1}.
\section{Dimensional analysis}
The formulas for entropy obtained above, namely, Eq.~\eq{e7}, \eq{mcv3}, \eq{mcv4}, and \eq{f2}, are functions of time with the same dimension. Now we introduce 
a dimensionless entropy:
\be{e6}
\mr\Delta \widetilde{S} = \frac{4}{ \rho L c_V} \(\frac{T_0}{\delta T}\)^2  \Delta S.
\ee
However, dimensionless time can be chosen differently depending  on the model of heat conduction.
It is well known that the description of physical processes does not depend on the choice of dimensions. Thus it is very useful to describe the process in dimensionless form. The description of heat conduction on different scales leads to different models of heat conduction. However, we would like to compare different properties of these models. Following Ref.~\cite{Glane_et_al._2018}, where the application of {\it Buckingham's $\pi$-theorem} was demonstrated in context with a problem coupling hydrodynamics and electrodynamics, we will apply this theorem to obtain dimensionless parameters, which describe the system above. We will construct and analyze a dimensional matrix $A$ as follows: The basic dimensions of the problem are time~-~$[T]$, length - $[L]$, mass~-~$[M]$ and temperature - $[\Theta]$. The brackets $[ \cdot ]$ shows the dimension of a quantity. Then the dimension matrix reads:
\be{dim1}
\begin{tabular}{ c c c c c c c }
	\hline
	& $T$ & $x$ & $t$ & $\alpha$ & $\tau$ & $c$   \\
	\hline
	$[L]$ & 0 & 1& 0 & 2 & 0 & 1   \\
	$[T]$  & 0 & 0& 1 &$-$1 & 1 &$-$1  \\
	$[\Theta]$ & 1 & 0& 0 & 0 & 0 & 0   \\
	\hline
	
\end{tabular}
\implies  A_{ij} = \begin{bmatrix}
	0 & 1& 0 & 2 & 0 & 1   \\
	0 & 0& 1 &-1 & 1 &-1  \\
	1 & 0& 0 & 0 & 0 & 0  
\end{bmatrix}_{ij}.
\ee
The components of this matrix are exponents of the dimensions. Each column gives the exponents for one considered physical quantity, {\it i.e.}, $[t] = [L]^0[T]^1[\Theta]^0[M]^0$.
The number $p$ of dimensionless quantities $\mathrm{\Pi}_i$, which is required to describe the system is the number $n$ of physical quantities  minus the rank $r$ of the matrix~$A$ . In the considered problem we have $n=6$, $r=3$. This means that the number $p$ of dimensions is $p = n - r = 3$. They are expressed as a product of physical quantities. $\mathrm{\Pi}_i = T^{l^i_1} x^{l^i_2} t^{l^i_3} \alpha^{l^i_4} \tau^{l^i_5} c^{l^i_6}$. As $\mathrm{\Pi}_i$ are dimensionless the following system of linear equations is obtained:
\be{dim2}
\sum_{j=1}^6 A_{ij}l_j^i = 0, \qquad i = 1,2,3.
\ee
The exponents $l^i_j$ are the components of the vectors from null space of the matrix~$A$ [the null space can be found by solving a system of linear Eqs. \eqref{dim2}]. Three linearly independent vectors are given by:
\ba{dim3}{c}
\mathbf{l}^1 = [ 0 \;-2 \;1 \;1 \;0 \;0]; \quad
\mathbf{l}^2 = [0\; 0\; -1\; 0\; 1\; 0]; \\
\mathbf{l}^3 = [ 0 \; -1 \; 1 \; 0 \; 0 \;1 ].
\ea
These vectors are needed to construct the following dimensionless quantities:
\be{dim4}
\mathrm{\Pi}_1 = \frac{t \alpha}{x^2}, \quad \mathrm{\Pi}_2 = \frac{\tau}{t}, \quad \mathrm{\Pi}_3 = \frac{c t}{x}.
\ee
By substituting dimensional coordinates $x =  x_{\mathrm{ref}} \widetilde{x}$, $t = t_{\mathrm{ref}} \widetilde{t} $, $T = T_{\mathrm{ref}} \widetilde{T} $ where tildes indicate dimensionless quantities. Then $\mathrm{\Pi}_i$ are redefined by means of reference scales:
\be{dim5}
\mathrm{\Pi}_1 = \frac{t_\mathrm{ref} \alpha}{x_{\mathrm{ref}}^2}, \quad  \quad \mathrm{\Pi}_2 = \frac{\tau}{t_{\mathrm{ref}}},  \quad \mathrm{\Pi}_3 = \frac{c t_{\mathrm{ref}}}{x_{\mathrm{ref}}}
.\ee
By using these dimensionless quantities our equations can be rewritten in the dimensionless form (see Table  \ref{table:1}).

\begin{table}[h]
	\centering
	\begin{tabular}{ c c c }
		\hline
		\hline \Top
		Ballistic & Hyperbolic & Classical \\ 
		\hline \Top
		$ \frac{\partial^2 \widetilde{T}}{\partial \widetilde{t}^2}+
		\frac{1}{\widetilde{t}}\frac{\partial \widetilde{T}}{\partial \widetilde{t}} = \Pi_3^2 \frac{\partial^2 \widetilde{T}}{ \partial \widetilde{x}^2}$ & $\Pi_2 \frac{\partial^2 \widetilde{T}}{\partial \widetilde{t}^2}+
		\frac{\partial \widetilde{T}}{\partial \widetilde{t}} = \Pi_1 \frac{\partial^2 \widetilde{T}}{ \partial \widetilde{x}^2}$ & $\frac{\widetilde{T}}{\partial \widetilde{t}} = \Pi_1 \frac{\partial^2 \widetilde{T}}{ \partial \widetilde{x}^2}$ \Bot \\ 
		\hline
		\hline
	\end{tabular}
	\caption{Dimensionless form of heat equations}
	\label{table:1}
\end{table}

The reference length scale $x_\mathrm{ref}$ of our problem is the wavelength of initial periodic perturbation, or we can choose the quantity inverse to wavenumber $k$: $x_\mathrm{ref} = 1/k$. Hence, our parameters will be the following ones:
\be{dim6}
\mathrm{\Pi}_1 = t_\mathrm{ref} \alpha k^2, \quad  \quad \mathrm{\Pi}_2 = \frac{\tau}{t_{\mathrm{ref}}},  \quad \mathrm{\Pi}_3 = c k t_{\mathrm{ref}}
.\ee
Now an appropriate timescale $t_\mathrm{ref}$ should be chosen. The reference timescale can be obtained in three different ways: $t_\mathrm{ref} = 1/\alpha k^2$  will be interpreted as the timescale of thermal diffusivity leading to $\mathrm{\Pi}_1=1$ and leaving $\mathrm{\Pi_2}$ and $\mathrm{\Pi_3}$ free; $t_\mathrm{ref} = \tau$ will be interpreted as the timescale of relaxation time leading to $\mathrm{\Pi}_2=1$ and leaving $\mathrm{\Pi_1}$ and $\mathrm{\Pi_3}$ free; $t_\mathrm{ref} = 1/ck$ will be interpreted as the time which a heat wave described by the ballistic heat Eq. \eq{4} needs to travel along one period of initial perturbation leading to $\mathrm{\Pi}_3=1$ and leaving $\mathrm{\Pi_1}$ and $\mathrm{\Pi_2}$ free.

\section{Results}
\begin{figure}[h]
	\caption{Plots of the entropy for ballistic heat equation calculated using CIT  for hyperbolic (dashed red line), CIT for ballistic heat equation (dashed black line),   EIT for hyperbolic (solid  red line), EIT for ballistic equation (solid black line) and CIT for classical heat equation (solid blue line). }
	\centering
	\includegraphics[width = 1\textwidth]{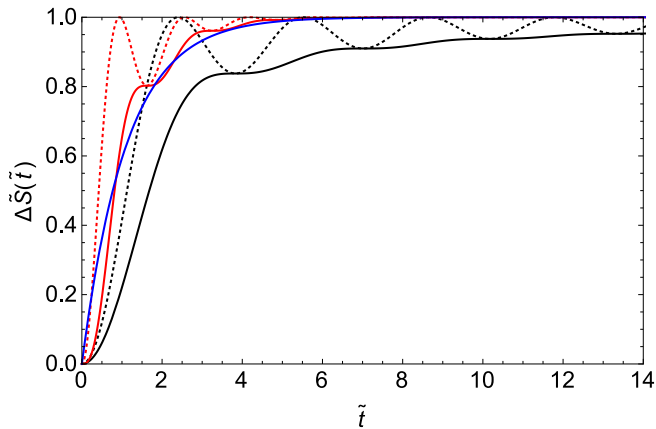}
	\label{Fig:CIT_EIT_compare}
\end{figure}

In this work three heat equations are considered: classical Eq. \eq{f1}, hyperbolic Eq. \eq{mcv1}, and the ballistic heat Eq. \eq{4}. By using {\it Buckingham's-$\pi$ theorem}, the system of these three equations can be rewritten in dimensionless form, see Table \ref{table:1}. By using the Eqs. \eqref{e7}, \eqref{mcv3}, \eqref{mcv4} and \eqref{f2} and the results obtained in the previous section we can obtain relations for dimensionless entropy.
A comparison of the three models is presented in Table \ref{tb1}, where $\widetilde{\omega} = \omega \tau$, $\widetilde{\omega}^* = \omega^* \tau$. Plots based on all three formulas with the timescale $t_\mathrm{ref} = 1/ck$ and the parameters $\mathrm{\Pi}_1 = 4$, $\mathrm{\Pi}_2 = 1$ are shown in Fig. \ref{Fig:CIT_EIT_compare}. Figures, similar to Fig. \ref{Fig:CIT_EIT_compare} can be found in Ref. \cite{Criado-Sancho_1993}. They contained the plots for the entropy, which was calculated with CIT and EIT for the hyperbolic heat Eq. \eq{mcv1}. In Fig. \ref{Fig:CIT_EIT_compare} we add new results for the ballistic heat Eq. \eq{4}.  It is seen that all three equations, the classic Eq. \eq{f1}, the hyperbolic Eq. \eq{mcv1}, and the ballistic Eq. \eq{4}, lead to same value of dimensionless entropy change at initial time and infinite time, no matter which approach we use, CIT or EIT.

\begin{table*}
	\begin{tabular}{ c c   c   c }
		\hline
		\hline \Top
		Equation & Ballistic & Hyperbolic & Classical \\ 
		\hline \Top
		Formula & $\ddot{T} + \frac{1}{t} \dot{T} = c^2T''$ & $\tau \ddot{T} + \dot{T} = \alpha T''$ & $\dot{T} = \alpha T''$ \Bot \\ 
		\hline 
		 \begin{tabular}{c}$\Delta \widetilde{S} $\\  CIT \end{tabular}  & $1-J_0^2(\widetilde{t})$ & \begin{tabular}{l} \Top $ 4 \alpha \tau  k^2   < 1:$ \\ $1 -e^{ -\widetilde{t} }\( A \sinh \widetilde{\omega} \widetilde{t} + \cosh \widetilde{\omega} \widetilde{t} \, \)^2,$ \\ $ 4 \alpha \tau  k^2   > 1:$ \\
			$1 - e^{-\widetilde{t}}\(A^* \sin \widetilde{\omega}^* \widetilde{t} + \cos \widetilde{\omega}^* \widetilde{t} \,\)^2$  \end{tabular} & $1-e^{-2 \widetilde{t}}$\\
		\hline
		\begin{tabular}{c} $\Delta \widetilde{S} $\\  EIT \end{tabular} & $1-J_0^2( \widetilde{t}) - J_1^2( \widetilde{t}) $ & 
		\begin{tabular}{l} $ \Top 4 \alpha \tau  k^2  < 1:$ \\ $1 -e^{ -\widetilde{t} }\{\( A \sinh \widetilde{\omega} \widetilde{t} + \cosh \widetilde{\omega} \widetilde{t} \, \)^2 - 4 \alpha \tau  k^2 A^2 \sinh^2 \widetilde{\omega} \widetilde{t} \},$ \\
			$ 4 \alpha \tau  k^2  > 1:$ \\
			$1 -e^{ -\widetilde{t} } \{\( A^* \sin \widetilde{\omega}^* \widetilde{t} + \cos \widetilde{\omega}^* \widetilde{t} \, \)^2 - 4 \alpha \tau  k^2 A^{*2} \sin^2 \widetilde{\omega}^* \widetilde{t}\},$
		\end{tabular}
		
		& --- \\
		
		\hline
		\Top
		Time scale & $\widetilde{t} = k c t$ & $\widetilde {t} = \frac{t}{\tau}$ & $\widetilde {t} = \alpha k^2 t$ \Bot \\
		\hline
		\Top
		Asymptotics & $1/\widetilde{t}$ 	& $e^{-\widetilde{t}}$ & $e^{-\widetilde{t}}$ \\
		\hline
		\hline
		
	\end{tabular}
	\caption{ Entropy change for sinusoidal initial temperature perturbation.}
	\label{tb1}
\end{table*}

Let us now consider the asymptotic behavior of the obtained entropies. An approximation formula for the Bessel function $J_n(t)$ at large times $t$, \mbox{is given by the following expression~\cite{Abramowitz_1964}:}
\be{d4}
J_n(t) \approx \sqrt{\frac{2}{\pi t}}\cos\(t - \frac{n\pi}{2} - \frac{\pi}{4}\) + \mathcal{O}\(\frac{1}{t}\)
.\ee
According to Eq. \eq{d4} the relations for the entropy for the ballistic heat equation from Eq. \eq{e7} reach the asymptotic value as a power law. Equations \eq{mcv3}, \eq{mcv4} and \eq{f2} indicate that the entropy for the hyperbolic Eq. \eq{mcv1} and classical Eq. \eq{f1} heat conduction will asymptotically be exponential. Also note that entropy values at initial time $0$ and at large times $ t \rightarrow \infty$ are given by $0$ and $1$ as seen from Fig. \ref{Fig:CIT_EIT_compare}.

\section{Conclusions}
In this work three equations of heat conduction were investigated, the classical Eq. \eq{f1}, the hyperbolic Eq. \eq{mcv1}, and the ballistic Eq. \eq{4}. The ballistic Eq.  \eq{4} of heat conduction was not considered from a phenomenological thermodynamics point of view before. This paper shows that the CIT approach can lead to a negative entropy production for the ballistic heat Eq. \eq{4}. Therefore the ballistic heat Eq. \eq{4} is considered within the framework of EIT. A general Eq. \eq{7} is obtained for calculating the entropy for Eq.\eq{4} with an arbitrary initial temperature perturbation. The example of an adiabatically closed system with an initial sinusoidal temperature perturbation is considered. For such a system a nonmonotonic increase of the total entropy indicates negative entropy production. This example shows that the CIT approach is not applicable for the ballistic Eq. \eq{4} and causes a negative entropy production and a nonmonotonic ``wavy" increase of the entropy. The entropy calculated with Eq. \eq{e4} obtained for EIT increases monotonically, and the entropy production stays nonnegative. 

We would like to note that in the case of CIT only the temperature  (as a function of $x$ and $t$) is needed to calculate the entropy [see Eq. \eq{i3}]. The expression for the temperature for a wide class of scalar lattices was obtained in Ref. \cite{Kuzkin_and_Krivtsov_2017}. Thus the results obtained in the current work within the framework of CIT can be extended to the case of scalar lattices.

Regarding the asymptotic behavior of entropy the following can be said:
There are papers showing that a harmonic system consisting of an infinite number of particles will approach spatial equilibrium for large times, see for example \cite{Spohn1977}. It is seen that at time $t \rightarrow \infty$, {\it i.e.}, when the system tends to equilibrium, the value of entropy is equal for all three models, no matter which approach is used, EIT or CIT. The change of dimensionless entropy $\Delta \widetilde{S}$ tends to one at large times. However, the asymptotic behavior of the entropy is significantly different. The entropy calculated for hyperbolic and for classical Fourier equations tends exponentially, $e^{-t}$, to an asymptotic value, whereas the entropy for ballistic heat Eq.~\eq{4} behaves according to a power law, $ 1/t$. 

If we now consider the problem from a discrete point of view, which in full generality is beyond the scope of this paper, then it can be said that the harmonic crystal can be decomposed into a system of independent modes and there will be no energy exchange between them. But in this system the irreversibility and the entropy rise is associated with the phases of oscillations, which distribute independently for long times---the details can be found in the pioneering work by  P.C.~Hemmer \cite{Hemmer_1959}.

We want to conclude by mentioning a frequently accepted concept of reversibility: Obviously the equation of ballistic heat Eq.~\eq{4}
\be{concl}
\ddot{T} + \frac{1}{t}\dot{T} = c^2 T''
\ee
is invariant with respect to time reversion: $t \rightarrow -t$. However, the results presented above show the irreversible nature of the ballistic heat conduction \eq{4} process, because of the increase of the total entropy of the system. Thus the mathematical time reversibility of the ballistic PDE \eq{4} is not correlated with its physical irreversibility.

%
\vspace{-10pt}

\begin{acknowledgments}
This work is supported by joint grant of the German Research Foundation (DFG) (Grant No. 405631704) and the Russian Science Foundation (Grant No.~19-41-04106). Contribution of W. H.~M\"uller and A. A. Sokolov is supported by the German Research Foundation, contribution of A. M.~Krivtsov and E. N.~Vilchevskaya is supported by the Russian Science Foundation.

The authors of this work thank Dmitry Korikov and Vitaly Kuzkin for the useful discussions. 
\end{acknowledgments}

\appendix
\setcounter{section}{-1}
\section{Solution for hyperbolic heat equation with sinusoidal initial temperature perturbation}
\label{appendixa}
In this section we present the solution of the hyperbolic heat Eq. \eq{mcv1} with the initial conditions Eq. \eq{init}. We find the solution in the form
\be{a1}
T(t,x) = \delta T f(t)\cos k x + T_0.
\ee
Substitution of Eq. \eq{a1} leads us to the following ODE for $f(t)$,
\be{a2}
\tau \ddot{f} + \dot{f} + \alpha k^2 f = 0,
\ee
with the following initial conditions:
\be{a3}
f(0) = 1,\qquad \dot{f}(0) = 0.
\ee
The solution is found in the form $e^{p t}$. The corresponding characteristic equation for the Eq. \eq{a2} is:
\be{a3a}
\tau p^2 + p + \alpha k^2 = 0.
\ee
Depending on the ratio of the parameters $\alpha$, $k$, $\tau$ this equation has real or complex roots:
\ba{a4}{l}
4 \alpha \tau k^2 < 1: \quad p_{1,2} =  \frac{-1 \pm \sqrt{1 - 4\alpha \tau k^2}}{2 \tau}, \\
4 \alpha \tau k^2 > 1: \quad p_{1,2} =  \frac{-1 \pm  i \sqrt{ 4\alpha \tau k^2-1}}{2 \tau}. \\
\ea
It leads us to the following fundamental solutions for Eq. \eq{a2}:
\ba{a5}{l}
4 \alpha \tau k^2 < 1: \\
f(t) = A_1 e^{\frac{-1 + \sqrt{1 - 4 \alpha \tau k^2}}{2 \tau}t} + A_2e^{\frac{-1 - \sqrt{1 - 4 \alpha \tau k^2}}{2 \tau}t}, \\
4 \alpha \tau k^2 > 1: \\
f(t) = e^{-\frac{t}{2 \tau} } \(B_1 \cos \frac{\sqrt{-1 + 4\alpha \tau k^2}}{2 \tau}t +  B_2 \sin \frac{\sqrt{-1 + 4\alpha \tau k ^ 2 }}{2 \tau}t \). \\
\ea
The coefficients $A_1$, $A_2$, $B_1$, $B_2$ are found by substitution of Eqs. \eq{a5} into the initial conditions Eq. \eq{a3}:
\ba{a6}{l l}
A_1 = \frac{1 + \sqrt{1 - 4 \alpha \tau k ^2}}{\sqrt{1 - 4 \alpha \tau k ^2}}, & A_2 = \frac{-1 + \sqrt{1 - 4 \alpha \tau k ^2}}{\sqrt{1 - 4 \alpha \tau k ^2}}, \\
B_1 = 1, & B_2 = \frac{1}{\sqrt{-1 + 4 \alpha \tau k^2}},
\ea
where $\alpha$ is the coefficient of thermal diffusivity. By using representations of hyperbolic trigonometric functions the solution for the temperature is:
\ba{a7}{ l }
4\alpha \tau k^2 < 1: \\ 
T(t,x) = \delta T e^{-\frac{t}{2\tau}} \( \cosh\omega t + A \sinh\omega t \) \cos k x + T_0,
 \\
4\alpha \tau k^2 > 1: \\ 
T(t,x) = \delta T e^{-\frac{t}{2 \tau}}\( \cos\omega^*t + A^* \sin\omega^*t \) \cos k x + T_0.
\ea
The solutions for the heat flux are obtained by substitution of Eqs. \eq{a7} into the second equation from Eq.~\eq{mcv1}:

\ba{a8}{l}
4\alpha \tau k^2 < 1: \\ 

\begin{aligned}
\tau \dot{h} + h = & \alpha \rho c_V k \delta T  e^{-\frac{t}{2\tau}} \( \cosh\omega t + A \sinh\omega t \) \sin k x,
\end{aligned} \\

4\alpha \tau k^2 > 1: \\ 

\begin{aligned}
\tau \dot{h} + h &=  \alpha \rho c_V k \delta T e^{-\frac{t}{2 \tau}}\( \cos\omega^*t + A^* \sin\omega^*t \) \sin k x.
\end{aligned}
\ea
The solution for Eqs. \eq{a8} is found in the form:
\be{a9}
h(t) = \alpha \rho c_V k \delta T f_h(t) \sin k x.
\ee
This leads us to the following inhomogeneous ODE's:
\ba{a10}{l}
4\alpha \tau k^2 < 1:  

\tau \dot{f_h} + f_h  =   e^{-\frac{t}{2\tau}} \( \cosh\omega t + A \sinh\omega t \) ,

 \\
4\alpha \tau k^2 > 1:

\tau \dot{f_h} + f_h = e^{-\frac{t}{2 \tau}}\( \cos\omega^*t + A^* \sin\omega^*t \),

\ea
with the initial condition 
\be{a11}
f_h(0) = 0.
\ee
The fundamental solutions for Eqs. \eq{a10} are:
\ba{a12}{l}
4\alpha \tau k^2 < 1:   f_h =  D_1 e^{-\frac{t}{\tau}} + 2 A  e^{-\frac{t}{2 \tau}} \sinh\omega t,  \\
4\alpha \tau k^2 > 1:    f_h =  D_2 e^{-\frac{t}{\tau}} + 2 A^*  e^{-\frac{t}{2 \tau}} \sin\omega^*t.
\ea
By substituting Eqs. \eq{a10} into the initial conditions Eq. \eq{a11} we obtain:
\be{a13}
D_1 = 0, \qquad D_2 = 0.
\ee
The solution for the heat flux is then:
\be{a14}
\begin{aligned}
&4\alpha \tau k^2 < 1:
h(t,x) =
2c_V \rho  \alpha k\delta T A e^{-\frac{t}{2 \tau}} \sinh \omega t \sin k x, \\
&4\alpha \tau k^2 > 1: 
h(t,x) = 2c_V \rho  \alpha k \delta T A^*  e^{-\frac{t}{2 \tau}} \sin\omega^*t \sin k x.
\end{aligned}
\ee
\bibliographystyle{apsrev4-1}
\bibliography{Sokolov_et_al_Entropy}

\end{document}